\documentclass[prl,aps,amstex,twocolumn,superscriptaddress,showpacs]{revtex4}

\usepackage{pxfonts}
\usepackage{graphicx}
\begin{document}

\title{Spin transport of indirect excitons}

\author{J.R. Leonard}
\author{Y. Kuznetsova}
\author{Sen Yang}
\author{L.V. Butov}
\affiliation{Department of Physics, University of California at San
Diego, La Jolla, California 92093-0319}

\author{T. Ostatnick\'y}
\affiliation{School of Physics and Astronomy, University of Southampton, SO17 1BJ, Southampton, United Kingdom}

\author{A. Kavokin}
\affiliation{Dipartimento di Fisica, Universita' "Tor Vergata" via della Ricerca
Scientifica, 00133 Rome, Italy}
\affiliation{School of Physics and Astronomy, University of Southampton, SO17 1BJ, Southampton, United Kingdom}

\author{A.C. Gossard}
\affiliation{Materials Department, University of California at Santa
Barbara, Santa Barbara, California 93106-5050}

\begin{abstract}
Spin transport of indirect excitons in GaAs/AlGaAs coupled quantum wells was observed by measuring the spatially resolved circular polarization of exciton emission. Exciton spin transport over several microns originates from a long spin relaxation time and long lifetime of indirect excitons.
\end{abstract}


\date{\today}

\maketitle

Spin physics in semiconductors includes a number of interesting
phenomena in electron transport, such as current-induced spin orientation (the spin Hall effect) \cite{Dyakonov1971,Hirsch1999,Sih2005}, spin-induced contribution to the current \cite{Dyakonov1971b}, spin injection \cite{Aronov1976}, and spin diffusion and drag
\cite{Kikkawa1999,Amico2001,Weber2005,Carter2006}. Besides the fundamental spin physics, there is also considerable interest in developing semiconductor electronic devices based on spin transport, which may offer advantages in dissipation, size and speed over charge-based devices, see \cite{Wolf2000,Awschalom2007} and references therein.

Optical methods have been used as a tool for precise probe and control of electron spin via photon polarization and, in particular, for studying electron spin transport in semiconductors \cite{Wolf2000,Stievater2001,Awschalom2007}. Excitons play a major role in the optical properties of quantum wells (QW) near the fundamental absorption edge. The spin dynamics of excitons in GaAs single QW was extensively studied in the past, see \cite{Maialle1993,Vinattieri1994} and references therein. It was found that the spin relaxation time of excitons in single QW is short, on the order of a few tens of ps. Because of the short spin relaxation time, no spin transport of excitons was observed in single GaAs QWs.

Here, we report on the observation of spin transport of indirect excitons in GaAs coupled quantum wells (CQW). The spin relaxation time of indirect excitons is orders of magnitude longer than for regular excitons. In combination with a long lifetime of indirect excitons, this makes possible spin transport of indirect excitons over several microns, which strongly exceeds the length scale of electron spin transport in metals.

The spin dynamics of excitons can be probed by the polarization resolved spectroscopy. In GaAs QW structures, the $\sigma^+$ ($\sigma^-$) polarized light propagating along the $z$-axis creates a heavy hole exciton with the electron spin state $s_z=-1/2$ ($s_z=+1/2$) and hole spin state $m_h=+3/2$ ($m_h=-3/2$). In turn, heavy hole excitons with $S_z=+1$ (-1) emit $\sigma^+$ ($\sigma^-$) polarized light. Excitons with $S_z=\pm 2$ are optically inactive. The polarization of the exciton emission
$P=(I_+ - I_-)/(I_+ + I_-)$ is determined by the recombination and spin relaxation processes. For an optically active heavy hole exciton, an electron or hole spin-flip transforms the exciton to an optically inactive state (Fig. 1a)
and do not cause the decay of the emission polarization. The polarization decays when both the electron and hole flip their spins. This can occur in the two-step process due to the separate electron and hole spin flips and the single-step process due to the exciton spin flip \cite{Andreani1990,Maialle1993,Vinattieri1994}. The rate equations describing these processes \cite{Maialle1993,Vinattieri1994} yield the polarization of the exciton emission $P=\tau_p/(\tau_p+\tau_r)$, where $\tau_p^{-1}=2(\tau_e+\tau_h)^{-1}+\tau_{ex}^{-1}$ is the relaxation
time of the emission polarization, $\tau_{ex}$ is the time for exciton
flipping between $S_z= \pm 1$ states, $\tau_e$ and $\tau_h$ are the
electron and hole spin flip times, and $\tau_r$ is the exciton recombination time (for the case when the splitting between $S_z=\pm 1$ and $\pm 2$ states is smaller than $k_BT$).

\begin{figure}
\begin{center}
\includegraphics[width=8.6cm]{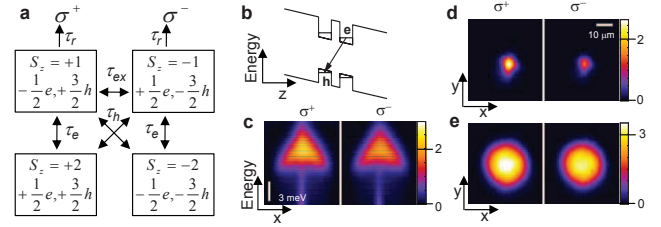}
\caption{\label{fig:fig1}(a) Exciton spin diagram, see text. (b) Energy diagram of the CQW structure; e, electron; h, hole. (c) Characteristic energy-$x$ images of the PL intensity of indirect excitons in $\sigma^+$ and $\sigma^-$ polarizations; $V_g=-1.1$ V, $E_{ex}=1.572$ eV, $P_{ex}=140\mu$W. $x-y$ images of the PL intensity of indirect excitons in $\sigma^+$ and $\sigma^-$ polarizations for (d) $P_{ex}=4.7\mu$W and (e) $P_{ex}=15\mu$W; $V_g=-1.1$ V, $E_{ex}=1.582$ eV.}
\end{center}
\end{figure}

In GaAs single QW, $\tau_h$ and $\tau_{ex}$ are typically in the range of tens of ps and are much shorter than $\tau_e$ and therefore $\tau_p \approx \tau_{ex}$ \cite{Uenoyama1990,Maialle1993,Vinattieri1994}. The short $\tau_{ex}$ results in a fast depolarization of the exciton emission within tens of ps in a single GaAs QW \cite{Maialle1993,Vinattieri1994}. However, $\tau_{ex}$ is determined by the strength of the exchange Coulomb interaction between the electron and hole. This gives an opportunity to control the depolarization rate by changing the electron-hole overlap, e.g. in QW structures with different QW widths or with an applied electric field \cite{Maialle1993,Vinattieri1994}.

The electron-hole overlap is drastically reduced in CQW structures where indirect excitons are composed of electrons and holes confined in separated wells (Fig. 1b). As a result of the small electron-hole overlap, the recombination time $\tau_r$ of indirect excitons is very long, typically in the range between tens of ns to tens of $\mu s$ \cite{Alexandrou1990,Zrenner1992}, orders of magnitude longer than $\tau_r$ in single QW, which is typically in the range of tens and hundreds of ps \cite{Maialle1993,Vinattieri1994}. Long lifetimes of indirect excitons make possible their transport over large distances, which can reach tens and hundreds of microns
\cite{Hagn1995,Larionov2000,Butov2002,Voros2005,Ivanov2006,Gartner2006}.
The small electron-hole overlap for indirect excitons should also result to a large $\tau_{ex} \propto \tau_r^2$ and in turn $\tau_p$, thus making possible exciton spin transport over substantial distances.

We probed exciton spin transport in a GaAs/AlGaAs CQW structure with two 8 nm GaAs QWs separated by a 4 nm Al$_{0.33}$Ga$_{0.67}$As barrier (see sample details in \cite{Butov1999} where the same sample was studied). The electric field across the sample was controlled by an applied gate voltage $V_g$. The excitons were photoexcited by a cw Ti:Sapphire laser tuned to the direct exciton energy, $E_{ex}=1.572$ eV, and focused to a spot $\sim 5 \mu$m in diameter. The spatial profile of the laser excitation spot was measured by the profile of the bulk GaAs emission from the excitation spot. The excitation was circularly polarized ($\sigma^+$). The emission images in $\sigma^+$ and $\sigma^-$ polarizations were taken by a CCD with an interference filter $800 \pm 5$ nm, which covers the spectral range of the indirect excitons. The spatial resolution was 1.4 micron. The spectra were measured using a spectrometer with resolution 0.3 meV. Characteristic x-energy spectra and x-y images are shown in Fig. 1c-e. Exciton density $n$ was estimated from the energy shift \cite{Ivanov2006} (for recent discussions of the interaction strength and density estimate see \cite{Schindler2008,Remeika2009}; the results on exciton spin transport reported here are practically insensitive to the interaction strength).

{\it Phenomenological model for exciton spin transport.} Rate equations combining the exciton spin relaxation equations \cite{Maialle1993,Vinattieri1994} with the drift-diffusion equation \cite{Ivanov2006} yield
\begin{eqnarray}
2\frac{dn_{\pm 1}}{dt}=&2\nabla\left[D\nabla n_{\pm 1} +\mu n_{\pm 1}\nabla\left(u_0n_b\right)\right]\nonumber\\
&-\frac{1}{\tau_r}n_{\pm 1}-\frac{1}{2\tau_p}\left(n_{\pm 1}-n_{\mp 1}\right)+\Lambda\delta_{+,\pm},
\end{eqnarray}
where \(D\) is the exciton diffusion coefficient, \(\mu=D/k_BT\) mobility, \(u_0\) interaction energy estimated by \(u_0=4\pi^2d/\epsilon\), \(n_b=n_{+1}+n_{-1}\), \(\tau_r\) radiative recombination time, \(\Lambda\) generation rate of +1 excitons, and \(\tau_p\) polarization relaxation time presented above. Both bright and dark exciton states are included in Eq. 1, however the fast hole spin flip process allowed the simplification by using \(\nabla n_{\pm1}=\nabla n_{\mp2}\) and \(\frac{\partial n_{\pm1}}{\partial t}=\frac{\partial n_{\mp2}}{\partial t}\). \(n_{+1}(r)\), \(n_{-1}(r)\), and $P(r)$ were calculated using Eq. 1 and compared to the experimental data. Note that the average polarization \(\left<P\right>=\left<\tau_P\right>/(\left<\tau_P\right>+\left<\tau_r\right>)\), which ignores the spatial dependence, allows a rough estimate of \(\tau_P\) without numerical simulations.

\begin{figure}
\begin{center}
\includegraphics[width=8.6cm]{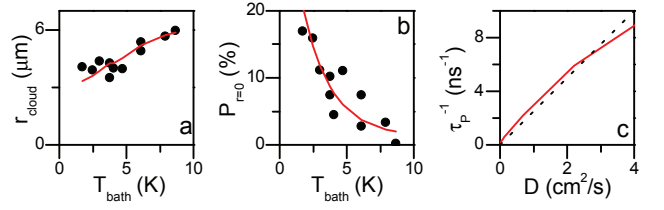}
\caption{\label{fig:fig2} Experimental (points) and simulated (curves) (a) exciton cloud radius and (b) degree of circular polarization at the center of the exciton cloud as a function of temperature. (c) The fit parameter \(1/\tau_P\) as a function of fit parameter \(D\). $V_g=-1.4$ V, $P_{ex}=10\mu$W, and $E_{ex}=1.572$ eV.}
\end{center}
\end{figure}

{\it Temperature dependence.} Increasing the temperature leads to the increase of the exciton cloud radius \(r_{cloud}\) and decrease of the circular polarization of exciton emission at the excitation spot center \(P_{r=0}\) (Fig. 2a,b). \(D\) and \(\tau_P\) were extracted from the measured \(r_{cloud}\), \(P_{r=0}\), and \(\tau_r\) \cite{Butov1999} via numerical simulations using Eq. (1). The obtained \(1/\tau_p\) vs. \(D\) is plotted in Fig 2c. The data show that (i) the depolarization time of the emission of indirect excitons reaches several ns, orders of magnitude longer than that of direct excitons in single QW \cite{Maialle1993,Vinattieri1994}, (ii) the polarization rapidly decreases with increasing temperature, and (iii) the decrease of polarization is correlated with the increase of diffusion coefficient.

{\it Density dependence.} Increasing the density leads to the increase of the exciton cloud radius \(r_{cloud}\) and decrease of the circular polarization of exciton emission at the excitation spot center \(P_{r=0}\) (Fig. 3a,b). At low densities, \(r_{cloud}\) is essentially equal to the excitation spot radius. The measured \(r_{cloud}\) and \(P_{r=0}\) were simulated using Eq. 1 with \(D\) and \(\tau_p\) as fitting parameters. The polarization degree of the exciton emission and the polarization relaxation time reduce with increasing density (Fig. 3b,d). (Note that no increase of $P$ with $n$ such as reported in Ref. \cite{Larionov2000} was observed in the present experiments.) Fig. 3e shows \(1/\tau_p\) as a function of \(D\) for the data in Fig. 3c,d.

\begin{figure}
\begin{center}
\includegraphics[width=8.6cm]{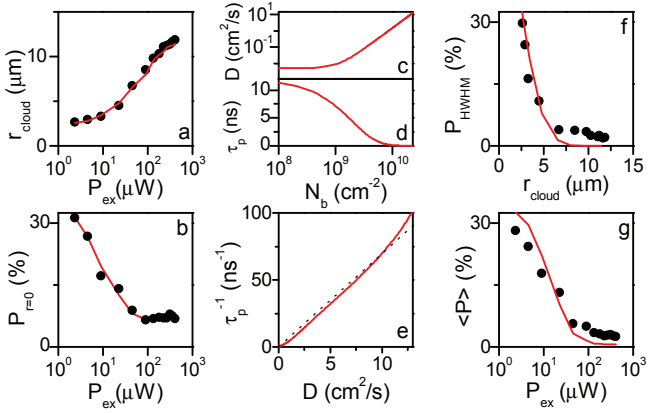}
\caption{\label{fig:fig3} Experimental (points) and simulated (curves) (a) exciton cloud radius \(r_{cloud}\) and (b) degree of circular polarization at the center of the exciton cloud \(P_{r=0}\) as a function of excitation density. (c,d) The fit parameters to \(r_{cloud}\) and \(P_{r=0}\) in (a,b) --- diffusion coefficient \(D\) and polarization relaxation time \(\tau_P\) as a function of \(n_b=n_{+1}+n_{-1}\). (e) \(1/\tau_P\) as a function of \(D\). (f) Experimental (points) and simulated (curve) polarization at HWHM of the exciton cloud \(P_{HWHM}\) as a function of \(r_{cloud}\). (g) Experimental (points) and simulated (curve) \(\left<P\right>\) as function of laser power. The simulations in (f) and (g) use the values of \(D(n)\) and \(\tau_P(n)\) obtained from fitting the data for \(r_{cloud}(n)\) and \(P_{r=0}(n)\) in (a,b).}
\end{center}
\end{figure}

{\it Spatial dependence: Exciton spin transport.} The polarization at HWHM of the exciton cloud $P_{HWHM}$ is observed up to several microns away from the origin (Fig. 3f). This gives a rough estimate for the length scale of exciton spin transport. Figure 3f,g also shows $P_{HWHM}$ and the spatially average polarization \(\left<P\right>\) calculated using Eq. 1 with \(D\) and \(\tau_p\) obtained from fitting \(r_{cloud}\) and \(P_{r=0}\) data in Fig. 3a,b and presented in Fig. 3c,d.

Straightforward characteristics of exciton spin transport is presented in Fig. 4. Figure 4a shows the measured PL in $\sigma^+$ and $\sigma^-$ polarization as a function of the distance from the excitation spot center $r$. Figure 4b shows the corresponding \(n_{+1}(r)\) and \(n_{-1}(r)\) calculated using Eq. 1 with \(D\) and \(\tau_p\) in Fig. 3c,d. The polarization profiles are wider than the excitation spot (Fig. 4c). This presents the spatial propagation of polarization, i.e. exciton spin transport. The measured and calculated data on exciton spin transport are in agreement (Figs. 3f, 4).

Note however that the calculated and measured data differ substantially at large $r$. This difference can originate from degrading experimental accuracy at large $r$ where the polarization is obtained by the division of two small quantities $I_+ - I_-$ and $I_+ + I_-$. However, there are also physical processes which can contribute to this difference. Note for instance that the above phenomenological model can be improved by including the effect of a higher exciton temperature at the excitation spot center \cite{Ivanov2006} and the spin Coulomb drag \cite{Amico2001,Weber2005}. The former should lead to a slower exciton spin relaxation at large $r$ where the exciton gas becomes colder, while the latter -- to reducing the exciton spin diffusion coefficient $D_s$ relative to $D$. Including these effects should make the modeling of exciton spin transport more accurate.

\begin{figure}
\begin{center}
\includegraphics[width=8.6cm]{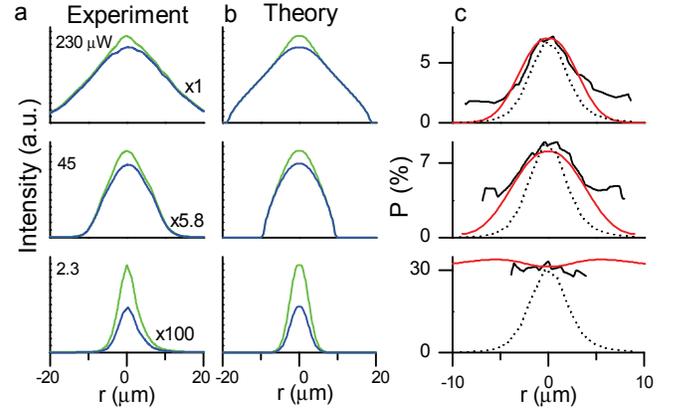}
\caption{\label{fig:fig3}
(a) Polarization resolved PL ($\sigma^+$, green; $\sigma^-$, blue) as a function of \(r\) for \(P_{ex}=\) 2.3, 45, and 230\(\mu W\) with estimated densities at \(r=0\) of \(9\cdot 10^8\), \(2\cdot 10^{10}\), and \(4\cdot 10^{10}cm^{-2}\) respectively. (b) Simulated $n_{+1}(r)$ and $n_{-1}(r)$ for the same exciton densities as in (a). (c) Experimental (black) and simulated (red) PL polarization as a function of \(r\) for the same exciton densities as in (a,b). The profile of the bulk emission,
which presents the excitation profile, is shown by a dotted line. $T_{bath}=1.7$ K. The simulations in (b) and (c) use the values of \(D(n)\) and \(\tau_P(n)\) obtained from fitting the data for \(r_{cloud}(n)\) and \(P_{r=0}(n)\) in Fig 3(a,b).}
\end{center}
\end{figure}

While a microscopic model of exciton spin transport is beyond the scope of this paper we briefly discuss the data below. First, we note that the parameters used in the calculations of exciton spin transport \(D\), \(\tau_P\), and \(\tau_r\) were obtained from other experiments: \(D\) -- from exciton transport, \(\tau_P\) -- from emission polarization at the excitation spot center, and \(\tau_r\) -- from PL kinetics. The agreement between the calculated and measured dependencies (Figs. 3f, 4) indicate that the major characteristics of exciton spin transport are determined by \(D\), \(\tau_P\), and \(\tau_r\). For \(\tau_r\), we just note here that it is long for indirect excitons, orders of magnitude longer than for regular direct excitons, and a long \(\tau_r\) is required for both exciton transport and spin transport over substantial distances since the (spin) diffusion length is limited by $\sqrt{D\tau_r}$ ($\sqrt{D_s\tau_r}$).

Next, we consider the effect of $D$ and \(\tau_p\) on exciton spin transport. Excitons are localized at low densities due to disorder and delocalized at high densities when the disorder is screened by repulsively interacting indirect excitons. Such localization-delocalization transition was studied earlier \cite{Butov2002,Ivanov2006,Remeika2009}. Localized excitons do not travel beyond the excitation spot while delocalized excitons spread over the distance $\sim \sqrt{D\tau_r}$. This accounts for the density dependence of $r_{cloud}$ and $D$ (Fig. 3a,c). $D$ also increases with temperature (Fig. 2a), which is consistent with thermal activation of indirect excitons over maxima of the disorder potential.

The ability to travel is required yet insufficient condition for spin transport. Beside a large $D$ and $\tau_r$, exciton spin transport over substantial distances also requires a long spin relaxation time. As highlighted in the introduction, for indirect excitons with a small electron-hole overlap both $\tau_r$ and $\tau_{ex}$ are long and, therefore, the polarization relaxation time \(\tau_p^{-1}=2(\tau_e+\tau_h)^{-1}+\tau_{ex}^{-1}\) is expected to be long. Indeed, $\tau_p$ for indirect excitons is very long at low temperatures and low densities reaching 10 ns (Fig. 2c, 3d), much longer than $\tau_p$ for regular excitons, which is in the range of tens of ps \cite{Maialle1993,Vinattieri1994}. However, $P$ and $\tau_p$ for indirect excitons drop with increasing temperature and density (Fig. 2b, 3b,d). For understanding this behavior we compare the variations of the polarization relaxation time and diffusion coefficient. Figure 2c and 3e show that $\tau_p^{-1}$ increases with $D$ both when the temperature or density is varied. This behavior is expected for the Dyakonov-Perel (DP) spin relaxation mechanism \cite{Dyakonov1971a} for which the spin relaxation time $\tau_{e,ex}^{-1}=\left<\Omega_{e,ex}^2 \tau\right>$, where \(\Omega_{e,ex}\) is the precession frequency caused by the energy splitting between the spin states, \(\tau \approx m_{ex}D/(k_BT)\) is the momentum scattering time, $m_{ex}$ is the exciton mass.

The measured dependence $\tau_p^{-1}(D)$ can be used for estimating the spin splitting. The data obtained with variable density (Fig. 3c) give a better accuracy for this estimate due to larger $r_{cloud}$ and, as a result, a higher accuracy in $D$. For indirect excitons with a very small electron-hole overlap, $\tau_{ex} \gg \tau_e$ and $\tau_p \approx \tau_e/2$ so that the polarization relaxation is determined by the electron spin relaxation. If the splitting of electron states is caused by the Dresselhaus mechanism \cite{Dresselhaus}, which is the most likely scenario, $\Omega_e=2\beta k/\hbar$ where \(k\) is the electron wave-vector. For the average thermal \(k\) of an electron in an exciton $k_T=\sqrt{2m_{ex}k_BT/\hbar^2}m_e/m_{ex}$, one obtains $\tau_p^{-1}=2\tau_e^{-1}=16\beta^2m_e^2D/\hbar^4$ and the measured $\tau_p^{-1}(D)$ (Fig. 3c) leads to the estimate of the spin splitting constant $\beta \approx 24 meV \AA$.

The value of $\beta$ can be roughly estimated from the bulk Dresselhaus constant GaAs $\gamma_c \approx 27.5 eV \AA^3$. In a 8~nm wide $\langle001\rangle$--oriented QW with infinitely high barriers $\beta=\gamma_c \langle k_z^2\rangle\approx \gamma_c (\pi/a)^2 \approx 42$meV$\AA$. More accurately, considering an electron in the
GaAs/AlGaAs CQW structure with a confining potential of 260~meV depth, we obtain $\beta\approx20\ \rm meV\AA$, in
agreement with the experiment.

The length scale for exciton spin transport reaches a few microns (Fig. 3f, 4c). It is large enough (i) for studying exciton spin transport by optical experiments, (ii) for studying spin-polarized exciton gases in microscopic patterned devices, e.g. in in-plane lattices \cite{Remeika2009}, in which the period can be below a micron, and (iii) for the development of spin-optoelectronic devices where {\it spin} fluxes of excitons can be controlled in analogy to the control of fluxes of unpolarized excitons in \cite{High2008} (the distance between source and drain in the excitonic transistor in \cite{High2008} was 3 $\mu m$; however, it is expected that the dimensions can be reduced below 1 $\mu m$ by using e-beam lithography).

In conclusion, the spatially resolved circular polarization of the
emission of long-life indirect excitons was studied. The polarization relaxation time of the emission of indirect excitons exceeds that of regular excitons by orders of magnitude and reaches ten ns. It reduces with increasing density and temperature in qualitative agreement with DP spin relaxation model. Spin transport of indirect excitons was observed. It originates from a long spin relaxation time and long lifetime of indirect excitons. The phenomenological model for exciton spin transport is in agreement with the experiment.

This work is supported by DOE Grant No. ER46449. T. O. and A. K. are supported by EPSRC.
We thank Misha Fogler, Lu Sham, and Congjun Wu for discussions.

\end{document}